\documentclass[preprint,aps,  nofootinbib]{revtex4}
\usepackage{epsf}
\usepackage{amscd}
\usepackage{amsmath}
\usepackage{epsfig}
\usepackage{amssymb}
\usepackage{amsfonts}
\usepackage{amsmath}
\usepackage{graphicx}
\usepackage{bm}

\begin{document}

\title{Model-Independent Results for SU(3) Violation in Twist-3 Light-Cone
Distribution Functions}
\author{Jiunn-Wei Chen\thanks{\texttt{jwc@phys.ntu.edu.tw}}, Hung-Ming Tsai%
\thanks{\texttt{r91222016@ntu.edu.tw}} and Ke-Chuan Weng\thanks{\texttt{%
r92222013@ntu.edu.tw}}}
\affiliation{Department of Physics, National Taiwan University, Taipei, Taiwan 10617}

\begin{abstract}
Using chiral symmetry we investigate the leading SU(3) violation in the
complete set of quark twist-3 light-cone distribution functions of the pion,
kaon, and eta, including the two-parton distributions $\phi _{M}^{p}$, $\phi
_{M}^{\sigma }$, and the three-parton distribution $\phi _{M}^{3}$. It is
shown that terms non-analytic in the quark masses do not affect the shape,
and only appear in the normalization constants. Predictive power is retained
including the leading analytic $m_{q}$ operators. With the symmetry
violating corrections we derive useful model-independent relations between $%
\phi _{\pi }^{p,\sigma ,3}$, $\phi _{\eta }^{p,\sigma ,3}$, $\phi
_{K^{+},K^{0}}^{p,\sigma ,3}$, and $\phi _{\bar{K}^{0},K^{-}}^{p,\sigma ,3}$%
. We also comment on the calculations of the moments of these distributions
using lattice QCD and light-cone sum rules.
\end{abstract}

\maketitle


\section{\protect\bigskip Introduction}

Meson light cone distribution functions (LCDFs) play important roles in high
energy hadronic exclusive processes \cite{bl,BBNS,HYCheng,KLS,SCET}. The
same LCDFs contribute in many processes relevant to measuring fundamental
parameters of the Standard Model~\cite{Bprocesses}, such as $B\rightarrow
\pi \ell \nu ,\eta \ell \nu $ which give the Cabibbo-Kobayashi-Maskawa (CKM) quark-mixing matrix element $|V_{ub}|$, $%
B\rightarrow D\pi $ used for tagging, and $B\rightarrow \pi \pi ,K\pi ,K\bar{%
K},\pi \eta ,\ldots $ which are important for measuring CP violation.

With the increasing accuracy in data from the $B$ factories, the
flavor dependence in LCDFs becomes important to understand the
flavor symmetry
breaking in processes like $B\rightarrow MM^{\prime }$ and $B\rightarrow MV$%
, where $M$ and $V$ are pseudoscalar and vector mesons, respectively. In
Ref. \cite{CS}, chiral perturbation theory\ (ChPT) was first applied to
study the leading SU(3) symmetry breaking effects in twist-2 LCDFs $\phi
_{M}^{P}(x)$. It was shown that terms non-analytic in the quark masses do
not affect the shape, and only appear in the normalization constants.
Furthermore, with the symmetry violating corrections useful
model-independent relations between $\phi _{\pi }^{P}$, $\phi _{\eta }^{P}$,
$\phi _{K^{+},K^{0}}^{P}$, and $\phi _{\bar{K}^{0},K^{-}}^{P}$\ were derived.

Recently ChPT has also been applied to the computation of hadronic twist-2
matrix elements~\cite{AS,CJ}. Many applications have been worked out,
e.g.,~chiral extrapolations of lattice data \cite%
{DMNRT,Qtwist2,Beane:2002vq,Detmold:2005pt}, generalized parton
distributions~\cite{Jq,piGPD,piGPD2}, large $N_{C}$ relations among nucleon
and $\Delta $-resonance distributions\cite{CJ}, soft pion productions in
deeply virtual Compton scattering~\cite{DVCSpi,DVCSpi-x,BS}, pion-photon
transition distributions \cite{pi-gam} and exclusive semileptonic B decays
\cite{Bdecays}. The method is also generalized to the multi-nucleon case
\cite{BS,CW}.

In this paper, we apply ChPT to higher twist matrix elements. We study the
leading SU(3) symmetry breaking in the complete set of quark twist-3 LCDFs,
including the two-parton distributions $\phi _{M}^{p}$, $\phi _{M}^{\sigma }$%
, and the three-parton distribution $\phi _{M}^{3}.$ Although those twist-3
contributions are parametrically suppressed by inverse powers of large
scales, they are numerically important in $B\rightarrow MM^{\prime }$, $%
B\rightarrow MV$ \cite{BBNS}\ and the meson electromagnetic form factor \cite%
{pi-ff}.

In the following sections, we will first summarize our main results on the
leading SU(3) symmetry breaking of LCDFs, then study the three twist-3 LCDFs
sequentially. Finally, we will comment on calculations of those quantities
using lattice QCD \cite{lattice0,lattice1,lattice2}\ and light cone sum
rules \cite{LCSR0,LCSR0.1,LCSR0.2,LCSR1,Ball:1998tj,LCSR2}.

\section{Summary of Results}

The two-parton LCDFs for pseudoscalar meson $M$ are defined by the matrix
element of the quark bilinear operator \cite{Braun1}
\begin{eqnarray}
\langle M^{b}(p)|\bar{q}_{\beta }^{\prime }(\frac{y}{2}n)\,[\frac{y}{2}n,-%
\frac{y}{2}n]\lambda ^{a}q_{\alpha }(-\frac{y}{2}n)|0\rangle &=&\frac{i}{4}%
\delta ^{ab}\int_{0}^{1}\!dx\,e^{i\left( x-1/2\right) \,yp\cdot n}\Big\{%
f_{M}^{P}p\!\!\slash\gamma _{5}\,\phi _{M}^{P}(x,\mu )  \notag \\
&&-\gamma _{5}\left( f_{M}^{p}\phi _{M}^{p}(x,\mu )-\frac{y}{6}f_{M}^{\sigma
}\sigma _{\mu \nu }\,p^{\mu }n^{\nu }\,\phi _{M}^{\sigma }(x,\mu )\right) %
\Big\}_{\alpha \beta }.\quad  \label{eq:2}
\end{eqnarray}%
where $n$\ is a constant light-like vector, $n^{2}=0$ and our octet matrices
are normalized so that $\mathrm{tr}[\lambda ^{a}\lambda ^{b}]=\delta ^{ab}$.
$[{\displaystyle {\frac{y }{2}}}n,-{\displaystyle {\frac{y }{2}}}n]$ denotes
the Wilson line connecting the quark bilinear located at different
space-time points on a light cone. $\phi _{M}^{P}$ is a twist-2 LCDF while $%
\phi _{M}^{p}$ and $\phi _{M}^{\sigma }$ are twist-3 LCDFs for pseudoscalar
meson $M$. $x$ is the quark momentum fraction and $\mu $ is the perturbative
QCD renormalization scale. For simplicity we work in the isospin limit and
the $\overline{\mathrm{MS}}$ scheme, and normalize the distributions so that
$\int \!\!dx\,\phi _{M}^{i}(x,\mu )\!=\!1$ with $i=P$, $p$, $\sigma $.

Generically from chiral symmetry the leading order SU(3) violation for $\phi
_{M}^{i}$ takes the form $[M\!=\!\pi ,K,\eta ]$
\begin{eqnarray}
\phi _{M}^{i}(x,\mu )\!\! &=&\!\!\phi _{0}^{i}(x,\mu
)+\!\!\!\!\!\sum\limits_{N=\pi ,K,\eta }\!\frac{m_{N}^{2}}{(4\pi f)^{2}}\Big[%
E_{M}^{N,i}(x,\mu )\ln \Big(\frac{m_{N}^{2}}{\mu _{\!\chi }^{2}}\Big)  \notag
\\
&&+F_{M}^{N,i}(x,\mu ,\mu _{\!\chi })\Big]\ .  \label{para}
\end{eqnarray}%
The functions $\phi _{0}^{i}$ , $E_{M}^{N,i}$, and $F_{M}^{N,i}$ are
independent of $m_{q}$, and are only functions of $\Lambda
_{\mathrm{QCD}}$, $\mu $, and $x$. $F_{M}^{N,i}$ also depends on the
ChPT dimensional regularization parameter $\mu _{\!\chi }$ which
cancels the $\ln (m_{N}^{2}/\mu _{\!\chi }^{2})$ dependence, so by
construction ${\phi _{M}^{i}}$ is $\mu _{\!\chi }$ independent.
Throughout the text, $\eta $ denotes the purely octet meson.

Using ChPT, we found very similar results to the twist-2 case \cite{CS} at
leading order in $SU(3)$ violation [$\mathcal{O}(m_{q})$] :

1) The twist-2 and twist-3 LCDFs are analytic in $m_{q}$, meaning that
\begin{equation}
E_{M}^{\pi ,i}(x)=0\,,\qquad E_{M}^{K,i}(x)=0\,,\qquad E_{M}^{\eta
,i}(x)=0\,.  \label{aa}
\end{equation}%
The leading logarithmic corrections can all be absorbed by the normalization
constants $f_{M}^{i}$.

2) By charge conjugation and isospin symmetry,
\begin{equation}
\phi _{\pi }^{i}(x)=\phi _{\pi }^{i}(1-x)\ ,\quad \phi _{\eta }^{i}(x)=\phi
_{\eta }^{i}(1-x)\ .  \label{A}
\end{equation}%
\ \ \ And by isospin symmetry
\begin{equation}
\phi _{K^{+}}^{i}(x)=\phi _{K^{0}}^{i}(x)\ ,\quad \phi _{K^{-}}^{i}(x)=\phi
_{\overline{K}^{0}}^{i}(x)\ .
\end{equation}
These two equations are true to all orders in $m_{q}$.

3)%
\begin{equation}
\phi _{K^{+}}^{i}(x)-\phi _{K^{+}}^{i}(1-x)=\phi _{K^{-}}^{i}(1-x)-\phi
_{K^{-}}^{i}(x)=\left( m_{s}-\overline{m}\right) \delta \phi ^{i}(x)\ ,
\label{C}
\end{equation}%
where $\delta \phi ^{i}(x)$ is $m_{q}$ independent.

4) Gell-Mann-Okubo-like relations exist among the octet mesons

\begin{equation}
\phi _{\pi }^{i}(x)+3\phi _{\eta }^{i}(x)=2[\phi _{K^{+}}^{i}(x)+\phi
_{K^{-}}^{i}(x)]\ .  \label{D}
\end{equation}

5) The three-parton LCDFs also have relations similar to 1)-4) [see Eqs.(\ref%
{ee}-\ref{eee})].

6) Statements 1)-5) are still true in quenched and partially quenched
simulations, and with the leading finite volume and finite lattice spacing
corrections.

7) The light cone sum rule results \cite{LCSR2} for twist-3 Gegenbauer
moments are consistent with the ChPT prediction $\left[ i=p,\sigma \right] $
\begin{equation}
4a_{2}^{K,i}=a_{2}^{\pi ,i}+3a_{2}^{\eta ,i}\text{ .}
\end{equation}%
The analogous ChPT relation for twist-2 moments puts a tight constraint on
the numerical values of $a_{2}^{M,P}$.

\section{\protect\bigskip Two-Parton LCDFs}

The operator product expansion of the non-local quark bilinear
operator in Eq.(\ref{eq:2}) gives rise to the twist-2 operator
$O_{k}^{P,a}$ and twist-3 operators $O_{k}^{p,a}$ and $\left(
O_{k}^{\sigma ,a}\right) ^{\mu \nu }$:
\begin{eqnarray}
O_{k}^{P,a} &=&\overline{\psi }\,n\!\!\!\slash\gamma _{5}\lambda ^{a}\left(
in\!\cdot \!\overleftrightarrow{D}\right) \,^{k}\,\psi  \label{eq:P} \\
O_{k}^{p,a} &=&\overline{\psi }\gamma _{5}\lambda ^{a}\left( in\cdot
\overleftrightarrow{D}\right) ^{k}\psi ,  \label{eq:p} \\
\left( O_{k}^{\sigma ,a}\right) ^{\mu \nu } &=&\overline{\psi }\sigma ^{\mu
\nu }\gamma _{5}\lambda ^{a}\left( in\cdot \overleftrightarrow{D}\right)
^{k+1}\psi ,  \label{eq:sig}
\end{eqnarray}%
where $\overleftrightarrow{D}=\overleftarrow{D}-\overrightarrow{D}$ and $%
k=0,1,2,\ldots $ Here having the vector indices dotted into $n^{\mu
_{0}}\cdots n^{\mu _{k(+1)}}$ has automatically projected onto the symmetric
and traceless part. The matrix elements of these operators yield%
\begin{eqnarray}
\langle M^{b}\left( p\right) |O_{k}^{P,a}|0\rangle &=&-if_{M}^{P}\delta
^{ab}(n\!\cdot \!p)^{k+1}\left\langle z^{k}\right\rangle _{M}^{P}\ ,
\label{eq:P-mat} \\
\left\langle M^{b}\left( p\right) |O_{k}^{p,a}|0\right\rangle
&=&-if_{M}^{p}\delta ^{ab}(n\cdot p)^{k}\left\langle z^{k}\right\rangle
_{M}^{p}\ ,  \label{eq:p-mat} \\
\left\langle M^{b}\left( p\right) |\left( O_{k}^{\sigma ,a}\right) ^{\mu \nu
}|0\right\rangle &=&-\frac{k+1}{3}f_{M}^{\sigma }\delta ^{ab}(n\cdot
p)^{k}\left( p^{\mu }n^{\nu }-p^{\nu }n^{\mu }\right) \left\langle
z^{k}\right\rangle _{M}^{\sigma }\ ,  \label{eq:sig-mat}
\end{eqnarray}%
where $z\equiv 1-2x$ and the moments are defined as
\begin{equation}
\left\langle z^{k}\right\rangle _{M}^{i}=\int_{0}^{1}dx(1-2x)^{k}\phi
_{M}^{i}(x)\ .
\end{equation}

Following similar procedures as in the twist-2 case \cite{CS}, we analyze
the twist-3 matrix elements.

\subsection{The $\protect\phi _{M}^{p}$ Analysis}

It is convenient to write
\begin{equation}
O_{k}^{p,a}=O_{k,LR}^{p,a}-O_{k,RL}^{p,a},
\end{equation}%
where
\begin{equation}
O_{k,LR}^{p,a}=\overline{\psi }_{L}\lambda _{LR}^{a}\big[in\!\cdot \!%
\overleftrightarrow{D}\big]^{k}\psi _{R},
\end{equation}%
and similarly for $O_{k,RL}^{p,a}$. $\psi _{L,R}=[(1\mp \gamma _{5})/2]\psi $
is the left(right)-handed quark field. The distinction between $\lambda
_{LR}^{a}$ and $\lambda _{RL}^{a}$ is only for bookkeeping purposes. We will
set $\lambda _{LR}^{a}=\lambda _{RL}^{a}=\lambda ^{a}$ at the end.

When $a=3$ or $8$, $O_{k}^{p,a}$ transforms simply under charge conjugation (%
$\mathcal{C}$), being even when $k$ is even, and odd when $k$ is odd. The
meson states $\pi ^{0}$ and $\eta $ (i.e., $M^{3,8}$) are $\mathcal{C}$
even. Thus from Eq.~(\ref{eq:p-mat}), $\left\langle z^{k}\right\rangle _{\pi
^{0},\eta }^{p}$ vanishes when $k$ is odd due to $\mathcal{C}$ (and using
isospin the same applies for $M=\pi ^{\pm }$). For all $a$'s the operator
would transform as
\begin{equation}
\mathcal{C}^{-1}O_{k}^{p,a}\mathcal{C}=(-1)^{k}O_{k}^{p,a}\,;
\end{equation}%
if we demanded that, under the $\mathcal{C}$ transformation
\begin{equation}
\lambda _{LR}^{a}\rightarrow \lambda _{LR}^{a\,T}\,,\qquad \lambda
_{RL}^{a}\rightarrow \lambda _{RL}^{a\,T}.  \label{lambda}
\end{equation}

To construct the corresponding hadronic ChPT operators, we define $\Sigma =%
\mathrm{exp}(2i\pi ^{a}\lambda ^{a}/f)$ and $m_{q}=\mathrm{diag}(\overline{m}%
,\overline{m},m_{S})=m_{q}^{\dag }$. Under a chiral $SU(3)_{L}\times
SU(3)_{R}$ transformation, we have
\begin{eqnarray}
\Sigma \rightarrow L\Sigma R^{\dag }, &\qquad &m_{q}\rightarrow
Lm_{q}R^{\dag } \\
\lambda _{LR}^{a}\rightarrow L\lambda _{LR}^{a}R^{\dag }, &\qquad &\lambda
_{RL}^{a}\rightarrow R\lambda _{RL}^{a}L^{\dag }.
\end{eqnarray}%
Under charge conjugation, $\Sigma \rightarrow \Sigma ^{T}$, $%
m_{q}\rightarrow m_{q}^{T}$, while $\lambda _{LR,RL}^{a}$ transform
according to Eq.~(\ref{lambda}). At next to leading order (NLO) in the $%
p^{2}/\Lambda _{\chi }^{2}$ and $m_{M}^{2}/\Lambda _{\chi }^{2}$ chiral
expansion,
\begin{equation}
O_{k}^{p,a}\rightarrow \sum_{i}c_{k,i}\mathcal{O}_{k,i}^{p,a}+\sum_{i}b_{k,i}%
\overline{\mathcal{O}}_{k,i}^{p,a},  \label{matching}
\end{equation}%
where the $\mathcal{O}$'s are the leading order (LO) and the $\overline{%
\mathcal{O}}$'s are NLO. The sum on $i$ runs over hadronic operators
which have the same transformation properties as $O_{k}^{p,a}$. The
ChPT Wilson coefficients $c_{k,i}$ and $b_{k,i}$ encode physics at
the scale $p^{2}\sim \Lambda _{\chi }^{2}$ and the operators encode
$p^{2}\ll \Lambda _{\chi }^{2} $.

At LO in the chiral expansion only one operator contributes in our analysis:
\begin{equation}
\mathcal{O}_{k,0}^{p,a}=\frac{f^{p}f}{8}\left[ 1+\left( -1\right) ^{k}\right]
\mathrm{Tr}\left[ \lambda _{LR}^{a}\Box ^{k}\Sigma ^{\dag }-\lambda
_{RL}^{a}\Box ^{k}\Sigma \right] ,  \label{eq:P-LO}
\end{equation}%
where $\Box ^{k}=(in\cdot \partial )^{k}$, $f^{p}=f_{M}^{p}$ in the chiral
limit. $\mathcal{O}_{k,0}^{p,a}$ vanishes when $k$ is odd due to charge
conjugation and SU(3) symmetry. The normalization of $\mathcal{O}%
_{k,0}^{p,a} $ is chosen such that $c_{0,0}=1$.

Note that the operator
\begin{equation}
\mathcal{O}_{k,1}^{p,a}=\frac{f^{p}f}{8}\left[ 1+\left( -1\right) ^{k}\right]
\mathrm{Tr}\left[ \lambda _{LR}^{a}\Sigma ^{\dagger }\left( \Box ^{k}\Sigma
\right) \Sigma ^{\dagger }-\lambda _{RL}^{a}\Sigma \left( \Box ^{k}\Sigma
^{\dagger }\right) \Sigma \right]
\end{equation}%
is also LO but not independent of $\mathcal{O}_{k,0}^{p,a}$. Since $\Box
^{k}(\Sigma ^{\dagger }\Sigma )=0$,
\begin{equation}
0=(\Box ^{k}\Sigma ^{\dagger })\Sigma +\Sigma ^{\dagger }(\Box ^{k}\Sigma
)+\ldots \,,  \label{eq:P-moveBox}
\end{equation}%
where the ellipse denotes $(\Box ^{k-m}\Sigma ^{\dagger })(\Box ^{m}\Sigma )$
terms that only contribute for matrix elements with more than one meson.
Thus, Eq. (\ref{eq:P-moveBox}) allows us to move factors of $\Box ^{k}$ onto
a neighboring $\Sigma $ and reduce $\mathcal{O}_{k,1}^{p,a}$ to $\mathcal{O}%
_{k,0}^{p,a}$.

At NLO there are two independent operators:
\begin{eqnarray}
\overline{\mathcal{O}}_{k,1}^{p,a} &=&\frac{f^{p}f}{8}\left[ 1+\left(
-1\right) ^{k}\right] \mathrm{Tr}\left[ m_{q}\Sigma ^{\dag }+\Sigma
m_{q}^{\dag }\right] \mathrm{Tr}\left[ \lambda _{LR}^{a}\Box ^{k}\Sigma
^{\dag }-\lambda _{RL}^{a}\Box ^{k}\Sigma \right] ,  \label{eq:p-NLO-1} \\
\overline{\mathcal{O}}_{k,2}^{p,a} &=&\frac{f^{p}f}{8}\mathrm{Tr}\Big[%
\lambda _{LR}^{a}\left\{ m_{q}^{\dag }\Sigma \Box ^{k}\Sigma ^{\dag
}+(-1)^{k}\left( \Box ^{k}\Sigma ^{\dag }\right) \Sigma m_{q}^{\dag }\right\}
\notag \\
&&-\lambda _{RL}^{a}\left\{ m_{q}\Sigma ^{\dag }\Box ^{k}\Sigma
+(-1)^{k}\left( \Box ^{k}\Sigma \right) \Sigma ^{\dag }m_{q}\right\} \Big].
\label{eq:p-NLO-2}
\end{eqnarray}%
All other NLO operators have derivatives on more than one $\Sigma $, or can
be reduced to $\overline{\mathcal{O}}_{k,1}^{p,a}$ and $\overline{\mathcal{O}%
}_{k,2}^{p,a}$ using the equations of motion. For instance, consider
\begin{eqnarray}
\overline{\mathcal{O}}_{k,3}^{p,a} &=&\frac{f^{p}f}{8}\mathrm{Tr}\Big[%
\lambda _{LR}^{a}\left\{ \Sigma ^{\dag }m_{q}\Box ^{k}\Sigma ^{\dag
}+(-1)^{k}\left( \Box ^{k}\Sigma ^{\dag }\right) m_{q}\Sigma ^{\dag }\right\}
\notag \\
&&-\lambda _{RL}^{a}\left\{ \Sigma m_{q}^{\dag }\Box ^{k}\Sigma
+(-1)^{k}\left( \Box ^{k}\Sigma \right) m_{q}^{\dag }\Sigma \right\} \Big].
\end{eqnarray}%
The sum and difference $\overline{\mathcal{O}}_{k,2}^{p,a}\pm \overline{%
\mathcal{O}}_{k,3}^{p,a}$ contain factors of $(\Sigma ^{\dagger }m_{q}\pm
m_{q}^{\dagger }\Sigma )$ and $(\Sigma m_{q}^{\dagger }\pm m_{q}\Sigma
^{\dagger })$. We can trade $\overline{\mathcal{O}}_{k,2}^{p,a}-\overline{%
\mathcal{O}}_{k,3}^{p,a}$ for operators with derivatives on more than one $%
\Sigma $ by using the equation of motion for $\Sigma $,
\begin{equation}
\Sigma ^{\dagger }(i\partial _{\mu })^{2}\Sigma =-(i\partial ^{\mu }\Sigma
^{\dagger })(i\partial _{\mu }\Sigma )\!+\!B_{0}(\Sigma ^{\dagger
}m_{q}\!-\!m_{q}^{\dagger }\Sigma )\,,  \label{eq:P-EOM-1}
\end{equation}%
together with the analogous equation for $\Sigma ^{\dagger }$. These
operators with derivatives on more than one $\Sigma $ do not generate
one-meson matrix elements at tree level and can be omitted from our
analysis. Thus only $\overline{\mathcal{O}}_{k,2}^{p,a}+\overline{\mathcal{O}%
}_{k,3}^{p,a}$ contributes and for simplicity we trade this for $\overline{%
\mathcal{O}}_{k,2}^{p,a}$. We can also insert more $\Sigma $ or $\Sigma
^{\dagger }$ fields into $\overline{\mathcal{O}}_{k,2}^{p,a}$ and get $%
\mathrm{Tr}\left[ \lambda _{LR}^{a}\Sigma ^{\dagger }m_{q}\Sigma
^{\dagger }\left( \Box ^{k}\Sigma \right) \Sigma ^{\dag }+\ldots
\right] $. But this
operator can be reduced to $\overline{\mathcal{O}}_{k,2}^{p,a}$ using Eq. (%
\ref{eq:P-moveBox}). Finally, we can consider operators where the power
suppression is generated by derivatives rather than a factor of $m_{q}$.
Boost invariance requires that these operators still have $k$ factors of $%
n^{\mu }$, so they will involve $\Box ^{k}$ just like the operators
we have been considering. To get power suppression with derivatives
we can either use $(\partial _{\mu }\Sigma ^{\dagger })(\partial
^{\mu }\Sigma )$ which has derivatives on more than one $\Sigma $,
or $\Sigma ^{\dagger }(\partial ^{\mu })^{2}\Sigma $ which can be
traded for operators with $m_{q}$'s using
Eq.~(\ref{eq:P-EOM-1}). Therefore, the operators with $m_{q}$'s in Eqs.~(\ref%
{eq:p-NLO-1}),(\ref{eq:p-NLO-2}) suffice.

\begin{figure}[t]
\vskip 0.cm
\centerline{
  \mbox{\epsfxsize=10.0truecm \hbox{\epsfbox{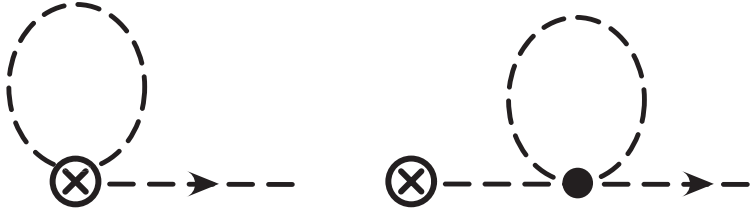}} }
  } \vskip-0.1cm \vspace{-0.5cm} 
\caption[1]{NLO loop diagrams, where here $\otimes $ denotes an insertion of
$\mathcal{O}_{k,0}^{p,a}$, and the dashed lines are meson fields.}
\label{Fig1}
\end{figure}

At NLO chiral logarithms can be obtained from loop diagrams involving the LO
operators as shown in Fig.\thinspace \ref{Fig1}. For $k=0$ the operator $%
O_{k=0}^{p,a}$ is the pseudoscalar current whose Fig.\thinspace \ref{Fig1}
graphs give the one-loop corrections to $f_{M}^{p}$. For odd $k$ the
one-loop graphs vanish due to the $\left[ 1+\left( -1\right) ^{k}\right] $
factor originated from $\mathcal{C}$. For any even $k>0$ the diagrams have a
term where all derivatives act on the outgoing meson line, and this gives
the same corrections as for $f_{M}^{p}$. The first diagram could have
additional contributions from derivatives acting inside the loop but it is
straightforward to show that these diagrams vanish identically since $%
n^{2}=0 $, and that the same holds true for LO operators with derivatives on
more than one $\Sigma $~\cite{AS}. Thus, we have shown that all possible
non-analytic corrections are contained in $f_{M}^{p}$ at NLO. This is true
for every moment, and so we conclude that the leading order SU(3) violation
of $\phi _{M}^{p}(x)$ is analytic in ${m_{q}}$.

At NLO the $\overline{\mathcal{O}}_{k,1}^{p,a}$, $\overline{\mathcal{O}}%
_{k,2}^{p,a}$ and the wave function renormalization counterterms all
contribute:
\begin{equation}
\langle M^{b}|c_{k,0}Z_{M}^{1/2}\mathcal{O}_{k,0}^{p,a}+\sum_{i=1}^{2}b_{k,i}%
\overline{\mathcal{O}}_{k,i}^{p,a}|0\rangle =-if_{M}^{p}\delta ^{ab}(n\cdot
p)^{k}\left\langle z^{k}\right\rangle _{M}^{p}\ .  \label{mat-1}
\end{equation}%
For $M=\pi $ and $K$, $f_{M}^{p}$ can be related to $f_{M}^{P}$ by relating $%
\overline{\psi }\gamma _{5}\lambda ^{a}\psi $ and the derivative of $%
\overline{\psi }\gamma ^{\mu }\gamma _{5}\lambda ^{a}\psi $ in the operator
level. Acting $D_{\mu }$ on both sides of
\begin{equation}
\langle M^{b}|\overline{\psi }(x)\gamma ^{\mu }\gamma _{5}\lambda ^{a}\psi
(x)|0\rangle =-if_{M}^{P}\delta ^{ab}p^{\mu }e^{ip\cdot x}\ ,
\end{equation}%
then by the equations of motion and Eq.(\ref{eq:P-mat}) \bigskip with $k=0,$
one obtains
\begin{eqnarray}
f_{\pi }^{p} &=&f_{\pi }^{P}\frac{m_{\pi }^{2}}{2\overline{m}}\text{ },
\notag \\
f_{K}^{p} &=&f_{K}^{P}\frac{m_{K}^{2}}{\overline{m}+m_{s}}\text{ }.
\label{f}
\end{eqnarray}%
These relations should be reproduced to all orders in ChPT. Indeed, direct
computations of the leading chiral logarithmic corrections of $f_{M}^{p}$, $%
f_{M}^{P}$ and $m_{M}^{2}$ confirm these results. There is no relation
between $f_{\eta }^{p}$ and $f_{\eta }^{P}$, however, because
\begin{equation}
iD_{\mu }\langle \eta |\overline{u}\gamma ^{\mu }\gamma _{5}u+\overline{d}%
\gamma ^{\mu }\gamma _{5}d-2\overline{s}\gamma ^{\mu }\gamma _{5}s|0\rangle
=-2\langle \eta |\overline{m}\left( \overline{u}\gamma _{5}u+\overline{d}%
\gamma _{5}d\right) -2m_{s}\overline{s}\gamma _{5}s|0\rangle \ ,  \label{eta}
\end{equation}%
which is not proportional to $\langle \eta |\overline{u}\gamma _{5}u+%
\overline{d}\gamma _{5}d-2\overline{s}\gamma _{5}s|0\rangle $ away from the
SU(3) limit.

By comparing the $k=0$ and $k\neq 0$ cases in Eq.(\ref{mat-1}), we obtain
\begin{eqnarray}
\left\langle z^{k}\right\rangle _{M}^{p} &=&\left\langle z^{k}\right\rangle
_{0}^{p}+2\Big\{\left( 1+(-1)^{k}\right) \mathrm{Tr}[m_{q}]\delta
^{ab}\left( b_{k,1}-b_{0,1}c_{k,0}\right)  \notag \\
&&+\mathrm{Tr}\left[ m_{q}\left( \lambda ^{b}\lambda ^{a}+(-1)^{k}\lambda
^{a}\lambda ^{b}\right) \right] \left( b_{k,2}-b_{0,2}c_{k,0}\left(
1+(-1)^{k}\right) /2\right) \Big\},  \label{TrM}
\end{eqnarray}%
where $\left\langle z^{k}\right\rangle _{0}^{p}=\left\langle
z^{k}\right\rangle _{M}^{p}$ in the chiral limit and $\left\langle
z^{0}\right\rangle _{M}^{p}=1$ is used.

For $k=2m+1$ (odd moments), the $\left\langle z^{k}\right\rangle _{M}^{p}$
structure yields
\begin{eqnarray}
\langle z^{2m+1}\rangle _{\pi }^{p} &=&\langle z^{2m+1}\rangle _{\eta
}^{p}=0,  \notag \\
\langle z^{2m+1}\rangle _{K^{+}}^{p} &=&\langle z^{2m+1}\rangle
_{K^{0}}^{p}=(m_{s}-\overline{m})b_{2m+1,2},  \label{eq:P-odd-moments} \\
\langle z^{2m+1}\rangle _{K^{-}}^{p} &=&\langle z^{2m+1}\rangle _{\overline{K%
}^{0}}^{p}=-(m_{s}-\overline{m})b_{2m+1,2}.  \notag
\end{eqnarray}%
For $k=2m$ (even moments), the $\left\langle z^{k}\right\rangle _{M}^{p}$
structure yields
\begin{eqnarray}
\langle z^{2m}\rangle _{\pi }^{p} &=&\left\langle z^{2m}\right\rangle
_{0}^{p}+2\overline{m}\alpha _{2m}^{p}+(2\overline{m}+m_{s})\beta _{2m}^{p},
\notag \\
\langle z^{2m}\rangle _{K}^{p} &=&\left\langle z^{2m}\right\rangle _{0}^{p}+(%
\overline{m}+m_{s})\alpha _{2m}^{p}+(2\overline{m}+m_{s})\beta _{2m}^{p},
\label{eq:P-even-moments} \\
\langle z^{2m}\rangle _{\eta }^{p} &=&\left\langle z^{2m}\right\rangle
_{0}^{p}+\frac{(2\overline{m}+4m_{s})}{3}\alpha _{2m}^{p}+(2\overline{m}%
+m_{s})\beta _{2m}^{p},  \notag
\end{eqnarray}%
where $\alpha _{2m}^{p}=(b_{2m,2}-b_{0,2}c_{2m,0})$ and $\beta
_{2m}^{p}=2(b_{2m,1}-b_{0,1}c_{2m,0})$. By isospin symmetry and charge
conjugation, the even moments of different pion states (or kaon states) are
equal. Eq.(\ref{eq:P-even-moments}) implies a Gell-Mann-Okubo-like relation:
\begin{equation}
\langle z^{2m}\rangle _{\pi }^{p}+3\langle z^{2m}\rangle _{\eta
}^{p}=4\langle z^{2m}\rangle _{K}^{p}.  \label{eq:P-G-O-1}
\end{equation}

The moment relations imply the LCDF relations in Eqs.(\ref{A}),(\ref{D}).
They also imply useful relations among the frequently used Gegenbauer
moments, defined by
\begin{equation}
a_{n}^{M,p}=\frac{2}{2n+1}\int_{0}^{1}\!\!\!\!dx\,C_{n}^{1/2}(2x\!-\!1)\phi
_{M}^{p}(x)\ ,
\end{equation}%
with $a_{0}^{M,p}=1$. Here $C_{n}^{1/2}(z)$ denote the Gegenbauer
polynomials which are even (odd) functions of $z$ when $n$ is even (odd).
\begin{equation}
\phi _{M}^{p}(x)=\sum\limits_{i=0}^{\infty
}a_{n}^{M,p}C_{n}^{1/2}(2x\!-\!1)\ .
\end{equation}%
Eqs.(\ref{eq:P-odd-moments})-(\ref{eq:P-G-O-1}) imply that
\begin{eqnarray}
4a_{2m}^{K,p} &=&a_{2m}^{\pi ,p}+3a_{2m}^{\eta ,p}\ ,\qquad  \label{a1} \\
a_{2m+1}^{\pi ,p} &=&a_{2m+1}^{\eta ,p}=0\,, \\
a_{2m+1}^{K^{0},p} &=&a_{2m+1}^{K^{+},p}=-a_{2m+1}^{\overline{K^{0}}%
,p}=-a_{2m+1}^{K^{-},p}\,.
\end{eqnarray}

\subsection{\protect\bigskip The $\protect\phi _{M}^{\protect\sigma }$
Analysis}

Following the analogous procedures, the twist-3 operator $(O_{k}^{\sigma
,a})^{\mu \nu }$ can be decomposed into
\begin{equation}
(O_{k}^{\sigma ,a})^{\mu \nu }=(O_{k}^{\sigma ,a})_{LR}^{\mu \nu
}-(O_{k}^{\sigma ,a})_{RL}^{\mu \nu }
\end{equation}%
with
\begin{equation}
(O_{k}^{\sigma ,a})_{LR}^{\mu \nu }=\overline{\psi }_{L}\sigma ^{\mu \nu
}\lambda _{LR}^{a}\big[in\!\cdot \!\overleftrightarrow{D}\big]^{k+1}\psi _{R}
\end{equation}%
and similarly for $(O_{k}^{\sigma ,a})_{RL}^{\mu \nu }$. Under charge
conjugation,
\begin{equation}
\mathcal{C}^{-1}(O_{k}^{\sigma ,a})^{\mu \nu }\mathcal{C}=(-1)^{k}(O_{k}^{%
\sigma ,a})^{\mu \nu }
\end{equation}%
if we demand $\lambda _{LR,RL}^{a}$ transform according to Eq.~(\ref{lambda}%
). At NLO in the chiral expansion, $O_{k}^{\sigma }$ is matched to LO and
NLO hadronic operators $\mathcal{O}_{k}^{\sigma }$ and $\overline{\mathcal{O}%
}_{k}^{\sigma}$:
\begin{equation}
(O_{k}^{\sigma ,a})^{\mu \nu }\rightarrow \sum_{i}c_{k,i}^{\sigma }(\mathcal{%
O}_{k,i}^{\sigma ,a})^{\mu \nu }+\sum_{i}b_{k,i}^{\sigma }(\overline{%
\mathcal{O}}_{k,i}^{\sigma ,a})^{\mu \nu }\ .  \label{matching2}
\end{equation}

Again, at LO there is only one hadronic operator
\begin{equation}
\left( \mathcal{O}_{k,0}^{\sigma ,a}\right) ^{\mu \nu }=\frac{f^{\sigma }f}{%
24}\left[ 1+\left( -1\right) ^{k}\right] \mathrm{Tr}\left[ \lambda
_{LR}^{a}\left( n^{[\nu }\partial ^{\mu ]}\Box ^{k}\Sigma ^{\dag }\right)
-\lambda _{RL}^{a}\left( n^{[\nu }\partial ^{\mu ]}\Box ^{k}\Sigma \right) %
\right] ,  \label{eq:T-LO}
\end{equation}%
where $n^{[\nu }\partial ^{\mu ]}=(n^{\nu }\partial ^{\mu }-n^{\mu }\partial
^{\nu })$. At NLO, there are two independent operators,
\begin{eqnarray}
\left( \overline{\mathcal{O}}_{k,1}^{\sigma ,a}\right) ^{\mu \nu } &=&\frac{%
f^{\sigma }f}{24}\left[ 1+\left( -1\right) ^{k}\right] \mathrm{Tr}\left[
m_{q}\Sigma ^{\dag }+\Sigma m_{q}^{\dag }\right] \mathrm{Tr}\left[ \lambda
_{LR}^{a}n^{[\nu }\partial ^{\mu ]}\Box ^{k}\Sigma ^{\dag }-\lambda
_{RL}^{a}n^{[\nu }\partial ^{\mu ]}\Box ^{k}\Sigma \right] ,
\label{eq:T-NLO-1} \\
\left( \overline{\mathcal{O}}_{k,2}^{\sigma ,a}\right) ^{\mu \nu } &=&\frac{%
f^{\sigma }f}{24}\mathrm{Tr}\Big[\lambda _{LR}^{a}\left\{ m_{q}^{\dag
}\Sigma n^{[\nu }\partial ^{\mu ]}\Box ^{k}\Sigma ^{\dag }+(-1)^{k}\left(
n^{[\nu }\partial ^{\mu ]}\Box ^{k}\Sigma ^{\dag }\right) \Sigma m_{q}^{\dag
}\right\}  \notag \\
&&-\lambda _{RL}^{a}\left\{ m_{q}\Sigma ^{\dag }n^{[\nu }\partial ^{\mu
]}\Box ^{k}\Sigma +(-1)^{k}\left( n^{[\nu }\partial ^{\mu ]}\Box ^{k}\Sigma
\right) \Sigma ^{\dag }m_{q}\right\} \Big].  \label{eq:T-NLO-2}
\end{eqnarray}%
These yield the following relations. For $k=2m+1$ (odd moments),

\begin{eqnarray}
\langle z^{2m+1}\rangle _{\pi }^{\sigma } &=&\langle z^{2m+1}\rangle _{\eta
}^{\sigma }=0,  \notag \\
\langle z^{2m+1}\rangle _{K^{+}}^{\sigma } &=&\langle z^{2m+1}\rangle
_{K^{0}}^{\sigma }=(m_{s}-\overline{m})b_{2m+1,2}^{\sigma },  \label{sig-odd}
\\
\langle z^{2m+1}\rangle _{K^{-}}^{\sigma } &=&\langle z^{2m+1}\rangle _{%
\overline{K}^{0}}^{\sigma }=-(m_{s}-\overline{m})b_{2m+1,2}^{\sigma }.
\notag
\end{eqnarray}%
For $k=2m$ (even moments),
\begin{eqnarray}
\langle z^{2m}\rangle _{\pi }^{\sigma } &=&\left\langle z^{2m}\right\rangle
_{0}^{\sigma }+2\overline{m}\alpha _{2m}^{\sigma }+(2\overline{m}%
+m_{s})\beta _{2m}^{\sigma },  \notag \\
\langle z^{2m}\rangle _{K}^{\sigma } &=&\left\langle z^{2m}\right\rangle
_{0}^{\sigma }+(\overline{m}+m_{s})\alpha _{2m}^{\sigma }+(2\overline{m}%
+m_{s})\beta _{2m}^{\sigma },  \label{sig-even} \\
\langle z^{2m}\rangle _{\eta }^{\sigma } &=&\left\langle z^{2m}\right\rangle
_{0}^{\sigma }+\frac{(2\overline{m}+4m_{s})}{3}\alpha _{2m}^{\sigma }+(2%
\overline{m}+m_{s})\beta _{2m}^{\sigma },  \notag
\end{eqnarray}%
where $\alpha _{2m}^{\sigma }=(b_{2m,2}^{\sigma }-b_{0,2}^{\sigma
}c_{2m,0}^{\sigma })$ and $\beta _{2m}^{\sigma }=2(b_{2m,1}^{\sigma
}-b_{0,1}^{\sigma }c_{2m,0}^{\sigma })$. These imply and%
\begin{equation}
\langle z^{2m}\rangle _{\pi }^{\sigma }+3\langle z^{2m}\rangle _{\eta
}^{\sigma }=4\langle z^{2m}\rangle _{K}^{\sigma }.
\end{equation}%
In terms of the Gegenbauer polynomials%
\begin{equation}
\phi _{M}^{\sigma }(x)=6x(1-x)\sum\limits_{i=0}^{\infty }a_{n}^{M,\sigma
}C_{n}^{3/2}(2x\!-\!1)\ ,
\end{equation}%
such that \bigskip the Gegenbauer moments are%
\begin{equation}
a_{n}^{M,\sigma }=\frac{4n+6}{6\!+\!9n\!+\!3n^{2}}\int_{0}^{1}\!\!\!\!dx%
\,C_{n}^{3/2}(2x\!-\!1)\phi _{M}^{\sigma }(x)\ ,  \label{g-moment}
\end{equation}%
with $a_{0}^{M,\sigma }=1$. Here $C_{n}^{3/2}(z)$ is an even (odd) function
of $z$ when $n$ is even (odd), thus we have
\begin{eqnarray}
4a_{2m}^{K,\sigma } &=&a_{2m}^{\pi ,\sigma }+3a_{2m}^{\eta ,\sigma }\ ,\qquad
\label{a2} \\
a_{2m+1}^{\pi ,\sigma } &=&a_{2m+1}^{\eta ,\sigma }=0\,, \\
a_{2m+1}^{K^{0},\sigma } &=&a_{2m+1}^{K^{+},\sigma }=-a_{2m+1}^{\overline{%
K^{0}},\sigma }=-a_{2m+1}^{K^{-},\sigma }\,.
\end{eqnarray}

The normalization $f_{M}^{\sigma }$ is related to $f_{M}^{P}$ for $M=\pi $
and $K$. By contracting
\begin{equation}
\left\langle M^{b}|\overline{\psi }(x)\sigma ^{\mu \nu }\gamma _{5}\lambda
^{a}i\overleftrightarrow{D}^{\alpha }\psi (x)|0\right\rangle =-\frac{1}{3}%
f_{M}^{\sigma }\delta ^{ab}\left( p^{\mu }g^{\nu \alpha }-p^{\nu }g^{\mu
\alpha }\right) e^{ip\cdot x}
\end{equation}%
with $g^{\nu \alpha }$ and making use of $D^{\mu }\langle M^{b}|\overline{%
\psi }(x)\gamma _{5}\lambda ^{a}\psi (x)|0\rangle =f_{M}^{p}\delta
^{ab}m_{M}^{2} p^\mu e^{ip\cdot x}$, one obtains%
\begin{eqnarray}
f_{\pi }^{\sigma } &=&f_{\pi }^{P}\left( \frac{m_{\pi }^{2}}{2\overline{m}}%
\right) \left[ 1-\left( \frac{2\overline{m}}{m_{\pi }}\right) ^{2}\right] \ ,
\notag \\
f_{K}^{\sigma } &=&f_{K}^{P}\left( \frac{m_{K}^{2}}{\overline{m}+m_{s}}%
\right) \left[ 1-\left( \frac{\overline{m}+m_{s}}{m_{K}}\right) ^{2}\right]
\ .  \label{ff}
\end{eqnarray}%
Direct computations of the chiral logarithms of $f_{\pi }^{\sigma }$ and $%
f_{K}^{\sigma }$ also confirm these results. Again, $f_{\eta }^{\sigma }$
and $f_{\eta }^{P}$ are not related unless in the SU(3) limit.

\section{\protect\bigskip Three Parton LCDFs}

The three-parton LCDF $\mathcal{T}_{M}$ for meson $M$ is defined by \cite%
{Braun1}:

\begin{eqnarray}
&&\langle 0|\bar{q}^{\prime }(yn)\,[yn,vyn]\sigma _{\mu \nu }\gamma
_{5}gG_{\sigma \rho }(vyn)\lambda ^{a}[vyn,-yn]q(-yn)|M^{b}(p)\rangle  \notag
\\
&=&if_{M}^{3}\delta ^{ab}\left[ p_{\sigma }p_{[\mu }g_{\nu ]\rho }-\left(
\sigma \leftrightarrow \rho \right) \right] T_{M}(v,yp\cdot n)\ ,
\end{eqnarray}%
where $g$ is the strong coupling constant and $G_{\sigma \rho }$ is the
gluon field strength tensor,
\begin{equation}
T_{M}(v,yp\cdot n)=\int d\underline{\alpha }e^{-iyp\cdot n\left( \alpha
^{\prime }-\alpha +v\alpha _{g}\right) }\mathcal{T}_{M}(\alpha ^{\prime
},\alpha ,\alpha _{g})\ ,
\end{equation}%
and%
\begin{equation}
\int d\underline{\alpha }=\int_{0}^{1}d\alpha d\alpha ^{\prime }d\alpha
_{g}\delta \left( 1-\alpha ^{\prime }-\alpha -\alpha _{g}\right) \ ,
\end{equation}%
and where $\alpha ^{\prime }$, $\alpha $ and $\alpha _{g}$ are momentum
fractions carried by $\bar{q}^{\prime }$, $q$ and the gluon. For simplicity,
we use the gauge $n\cdot A=0$, such that the Wilson lines $[yn,vyn]=$ $%
[vyn,-yn]=1$. After the operator product expansion, the resulting twist-3
operator is
\begin{equation}
O_{k,j}^{3,a}=\sum\limits_{i=0}^{k}\bar{q}^{\prime }\left( in\cdot
\overleftarrow{\partial }\right) ^{i}\sigma _{\mu \nu }\gamma _{5}g\left[
\left( in\cdot \overrightarrow{\partial }\right) ^{j}G_{\sigma \rho }\right]
\lambda ^{a}\left( in\cdot \overrightarrow{\partial }\right) ^{k-i}q\ ,
\end{equation}%
where we have suppressed the $\mu $, $\nu $, $\sigma $, and $\rho $ indexes
on the left hand side. Note that $O_{k,j}^{3,a}$ transforms in the same way
as $(O_{k}^{\sigma ,a})^{\mu \nu }$ under chiral transformation and
\begin{equation}
\mathcal{C}^{-1}O_{k,j}^{3,a}\mathcal{C}=(-1)^{k}O_{k,j}^{3,a}\,.
\end{equation}%
Thus the matrix element
\begin{equation}
\langle 0|O_{k}^{3,a}|M^{b}\rangle =if_{M}^{3}\delta ^{ab}\left[ p_{\sigma
}p_{[\mu }g_{\nu ]\rho }-\left( \sigma \leftrightarrow \rho \right) \right]
\left( n\cdot p\right) ^{k+j}\int d\underline{\alpha }\left( \alpha ^{\prime
}-\alpha \right) ^{k}\alpha _{g}^{j}\mathcal{T}(\alpha ^{\prime },\alpha
,\alpha _{g})
\end{equation}%
implies%
\begin{equation}
\mathcal{T}_{\pi \left( \eta \right) }(\alpha ^{\prime },\alpha ,\alpha
_{g})=\mathcal{T}_{\pi \left( \eta \right) }(\alpha ,\alpha ^{\prime
},\alpha _{g})\ ,  \label{ee}
\end{equation}%
\begin{equation}
\mathcal{T}_{K^{+}}(\alpha ^{\prime },\alpha ,\alpha _{g})-\mathcal{T}%
_{K^{+}}(\alpha ,\alpha ^{\prime },\alpha _{g})=-\mathcal{T}_{K^{-}}(\alpha
^{\prime },\alpha ,\alpha _{g})+\mathcal{T}_{K^{-}}(\alpha ,\alpha ^{\prime
},\alpha _{g})\propto \left( m_{s}-\overline{m}\right) ,
\end{equation}%
\begin{equation}
2\left[ \mathcal{T}_{K^{+}}(\alpha ^{\prime },\alpha ,\alpha _{g})+\mathcal{T%
}_{K^{-}}(\alpha ^{\prime },\alpha ,\alpha _{g})\right] =\mathcal{T}_{\pi
}(\alpha ^{\prime },\alpha ,\alpha _{g})+3\mathcal{T}_{\eta }(\alpha
^{\prime },\alpha ,\alpha _{g})\ ,  \label{eee}
\end{equation}%
together with $\mathcal{T}_{K^{+}}=\mathcal{T}_{K^{0}}$ and $\mathcal{T}%
_{K^{-}}=\mathcal{T}_{\overline{K}^{0}}$ by isospin. Eq.(\ref{ee}) was also
obtained in Refs. \cite{Braun1,Pirjol}.

The fact that $O_{0,j}^{3,a}$, $(O_{0}^{\sigma ,a})^{\mu \nu }$ and $%
O_{0}^{p,a}$ transform in the same way under chiral transformation implies
that the normalization constants $f_{M}^{3}$, $f_{M}^{\sigma }$ and $%
f_{M}^{p}$ receive the same leading logarithmic corrections. This is true
between $f_{M}^{p}$ and $f_{M}^{\sigma }$ in Eqs. (\ref{f}) and (\ref{ff})
due to the similar structures of the corresponding LO operators in Eqs. (\ref%
{eq:P-LO}) and (\ref{eq:T-LO}). Also, at NLO the ratio%
\begin{equation}
R_{M}=\frac{f_{M}^{3}}{f_{M}^{\sigma }}
\end{equation}%
satisfies%
\begin{equation}
R_{K^{+}}-R_{K^{-}}\propto \left( m_{s}-\overline{m}\right) \ ,
\end{equation}%
and%
\begin{equation}
4R_{K}=R_{\pi }+3R_{\eta }\ .
\end{equation}%
Analogous relations also exist for the ratio $f_{M}^{\sigma }/f_{M}^{p}$ at
NLO.

As pointed out in Ref. \cite{Braun1} (see also \cite{LCSR1,Pirjol}), the
three twist-3 LCDFs are related by the equations of motion. This implies in
ChPT, the counterterms associated with the three LCDFs are also related. It
can be shown straightforwardly that the SU(3) relations presented in this
work are consistent with the constraints imposed by the equations of motion.

\section{Lattice QCD Chiral Extrapolations}

The first few moments of the twist-2 and twist-3 LCDFs can be calculated
using lattice QCD. Currently only the second twist-2 moment of pion $%
\left\langle z^{2}\right\rangle _{\pi }^{P}$ is computed \cite%
{lattice0,lattice1,lattice2}. The latest results on $\left\langle
z^{2}\right\rangle _{\pi }^{P}$ have converged to $\left\langle
z^{2}\right\rangle ^{\overline{MS}}(\mu =2.67$ GeV$%
^{2})=0.280(49)_{-0.013}^{+0.030}$ \cite{lattice1} and $\left\langle
z^{2}\right\rangle ^{\overline{MS}}(\mu ^{2}=5$ GeV$^{2})=0.281(28)$ \cite%
{lattice2} with the lightest $m_{\pi }$ to be $490$ MeV and $550$ MeV,
respectively. In these two calculations, and many others, chiral
extrapolations to the physical point $m_{\pi }=140$ MeV were performed. Thus
the model-independent results on the $m_{q}$ dependence of LCDFs obtained in
Ref. \cite{CS} and this work are valuable to reduce the systematic errors in
the extrapolations.

A very interesting result we found both in the twist-2 and twist-3 LCDFs is
that, in the continuum, all the moments of those distributions are analytic
in $m_{q}$ at $\mathcal{O}(m_{q})$. Remarkably, this nice feature is
retained in a lattice theory with (partially) quenched, finite volume and
the leading finite lattice spacing effects included. It is easy to see why.
In all the cases, the leading logarithms are generated by one loop diagrams
with insertion of the LO $\mathcal{O}_{k,0}$ operator [with the trace
replaced by super-trace when (partially) quenched]. These diagrams correct $%
c_{k,0}$ by a universal ($k$ independent) logarithm which can then be
absorbed into the normalization constant. Thus the moments are analytic in $%
m_{q}$. This feature, however, is only true at $\mathcal{O}(m_{q})$. At $\mathcal{O}(m_{q}^{2})$%
, the $m_{q}^{2}\ln ^{2}m_{q}$\ corrections to $c_{k,0}$\ are
generated by two loop diagrams with insertion of the LO
$\mathcal{O}_{k,0}$\ operator. These corrections are $k$\
independent and can be absorbed into the normalization factors. The
$m_{q}^{2}\ln m_{q}$\ corrections, however, can arise from one-loop
diagrams with insertion of the NLO $\overline{\mathcal{O}}_{k,1}$\
and $\overline{\mathcal{O}}_{k,2}$\ operators which depend on
$b_{k,1}$\ and
$b_{k,2}$\ and correct $c_{k,0}$\ in a $k$\ dependent way, thus the $%
m_{q}^{2}\ln m_{q}$\ corrections cannot be absorbed by the normalization
constants. Generalizing the above argument to $\mathcal{O}(m_{q}^{i})$, the $%
m_{q}^{i}\ln ^{i}m_{q}$\ corrections can be absorbed by the
normalization constants, thus\ asymptotically, the LCDFs has the
quark mass dependence
\footnote{%
Consider a generic diagram giving an $\mathcal{O}(\epsilon ^{d})$
contribution to the twist-2 matrix element $\langle M(p)|O_{k}^{P}|0\rangle $%
, where $\epsilon \sim p\sim m_{\pi }$. If the diagram has a $\mathcal{O}%
(\epsilon ^{k+1+2n})$ pionic operator insertion, $I$ internal propagators, $L
$ loops, and $V_{i}$ vertex of order $\epsilon ^{d_{i}}$, then $%
d=k+1+2n-2I+4L+\sum_{i}V_{i}d_{i}$. After using the topological identity $%
L=I-\sum_{i}V_{i}$ and removing the $\epsilon ^{k+1}$ kinematical factor,
this diagram gives a $\mathcal{O}(\epsilon ^{2h})$ contribution to the
moment $\left\langle z^{k}\right\rangle _{M}^{P}$ with $h=n+L+\sum_{i}V_{i}%
\left( d_{i}/2-1\right) $. Since $n\geq 0$ and $d_{i}\geq 2$, $h\geq L$. In
general, the diagram has the contribution $\sum_{i=0}^{L}b_{i}m_{q}^{h}\ln
^{i}m_{q}$, with the power of logarithm not bigger than $L$ and $%
m_{q}\propto \sqrt{m_{\pi }}=\mathcal{O}(\sqrt{\epsilon })$. The $%
m_{q}^{h}\ln ^{h}m_{q}$ term can only come from the diagram with an
insertion of the LO pionic operator ($n=0$) and can be absorbed into the
normalization. The arguments can be easily generalized to the twist-3 cases.}
\begin{equation}
a_{00}+\sum\limits_{i=1}^{\infty }\sum\limits_{j=0}^{i-1}a_{ij}m_{q}^{i}\ln
^{j}m_{q}\ ,
\end{equation}%
where $a_{00}$ and $a_{ij}$\ are independent of $m_{q}$.

For different types of lattice simulations, the leading chiral corrections
from the one loop diagrams can still be absorbed into the normalization
constants, such that the LCDFs remain analytic at NLO. Different types of
simulations give different normalization constants and different analytic
contributions to the LCDFs. For example, the $(2\overline{m}+m_{s})$ term in
Eq.(\ref{eq:P-even-moments}) is originated from $\mathrm{Tr}[m_{q}]$ in Eq.(%
\ref{TrM}). In the (partially) quenched theory, $\mathrm{Tr}[m_{q}]$
is
replaced by $\mathrm{Str}[m_{q}]$ which is zero in the quenched theory \cite%
{Quenched} and is $(2\overline{m}_{sea}+m_{s,sea})$ in the partially
quenched \cite{PQ}\ theory, with $m_{q,sea}$ denoting sea quark
masses. It is easy to check that Eq.(\ref{eq:P-G-O-1}) still holds
in spite of (partially) quenching. As for finite volume effects, the
counterterm structures are the same as those in the infinite volume
calculations since the effects are
infrared in origin. Thus in simulations with twisted \cite%
{Luscher,Paolo,twistedx}\ or partially twisted \cite{PT1,PT2}\
boundary conditions for fermions, the leading finite volume effects
are absorbed by the normalization constant and the moments are all
identical to those in the infinite volume calculations. For lattice
fermions whose chiral symmetry is broken by finite lattice spacing
$a$, new counterterms in powers of $a$ need to be added to account
for the leading finite $a$ effects. Since these
counterterms are SU(3) symmetric, Gell-Mann-Okubo-like relations like Eq.(%
\ref{eq:P-G-O-1}) will still be true. The same is true for mixed action
simulations with chiral valance quarks but Wilson or staggered sea quarks.

\section{Comparison with Light Cone Sum Rule Results}

The first few Gegenbauer moments for the octet mesons have been computed
using light cone sum rules \cite%
{LCSR0,LCSR0.1,LCSR0.2,LCSR1,Ball:1998tj,LCSR2}. The twist-2 moments $%
a_{2}^{M,P}$ (defined in the same way as $a_{2}^{M,\sigma }$ in Eq.(\ref%
{g-moment})) were determined in different references. Quite different values
for $a_{2}^{\pi ,P}(\mu =1$ GeV$)$, $0.44$ \cite{LCSR0} and $0.20$-$0.25$
\cite{LCSR0.1}, are obtained using light cone sum rules. After combining
these values with various constraints from experimental data \cite%
{data,datax}, a reasonable range is assigned \cite{datax}:
\begin{equation}
0\leq a_{2}^{\pi ,P}(1\text{ GeV})\leq 0.3\ .  \label{bbb}
\end{equation}%
For moments of $K$ and $\eta $, light cone sum rules of Refs. \cite{LCSR2}
and \cite{LCSR1} yield
\begin{equation}
a_{2}^{K,P}(1\text{ GeV})=0.16\ ,\quad a_{2}^{\eta ,P}(1\text{ GeV})=0.2\ ,
\label{bb}
\end{equation}%
respectively. Since the range of $a_{2}^{\pi ,P}$ is big, the results in
Eqs.(\ref{bbb}),(\ref{bb}) are still consistent with the ChPT relation found
in Ref. \cite{CS};
\begin{equation}
\rho =\frac{a_{2}^{\pi ,P}+3a_{2}^{\eta ,P}}{4a_{2}^{K,P}}=1+\mathcal{O}%
(m_{q}^{2})\ .  \label{rho}
\end{equation}%
However, this equation could provide a tight constraint among $a_{2}^{M,P}$
since one expects the $\mathcal{O}(m_{q}^{2})$ correction to be 10\% at
most. \

As for $a_{1}^{K^{-},P}$, recent results yield $a_{1}^{K^{-},P}=0.05\pm 0.02$
\cite{Khodjamirian:2004ga}, $0.10\pm 0.12$ \cite{Braun:2004vf} and $0.050\pm
0.025$ \cite{Ball:2005vx}. The positive sign corresponds to the $s$-quark in
a $K^{-}$ meson carrying a bigger momentum fraction than the $\overline{u}$%
-quark.

In Ref. \cite{LCSR2}, twist-3 Gegenbauer moments are computed (see also \cite%
{Huang}):
\begin{eqnarray}
a_{2}^{M,p} &=&30\eta _{3}^{M}-\frac{5}{2}\rho _{M}^{2}\ ,  \notag \\
a_{4}^{M,p} &=&-3\eta _{3}^{M}\omega _{3}^{M}-\frac{27}{20}\rho _{M}^{2}-%
\frac{81}{80}\rho _{M}^{2}a_{2}^{M,P}\ ,  \label{cc} \\
a_{2}^{M,\sigma } &=&5\eta _{3}^{M}-\frac{1}{2}\eta _{3}^{M}\omega _{3}^{M}-%
\frac{7}{20}\rho _{M}^{2}-\frac{3}{5}\rho _{M}^{2}a_{2}^{M,P}\ .  \notag
\end{eqnarray}%
The numerical values of $\eta _{3}^{M}$ and $\omega _{3}^{M}$ are
independent of $M,$ while $\rho _{M}^{2}\propto m_{M}^{2}=\mathcal{O}(m_{q})$
and $a_{2}^{M,P}=a_{2}^{P}(1+\mathcal{O}(m_{q}))$ with $%
a_{2}^{P}=a_{2}^{M,P} $ in the chiral limit. Eq.(\ref{cc}) satisfies the
relations Eqs.(\ref{a1}),(\ref{a2}) and, hence, is consistent with chiral
symmetry.

\section{Conclusions}

Using chiral symmetry we have investigated the leading SU(3) violation in
the complete set of twist-3 light-cone distribution functions of the pion,
kaon, and eta, including the two-parton distributions $\phi _{M}^{p}$, $\phi
_{M}^{\sigma }$ and the three-parton distribution $\phi _{M}^{3}$. It has
been shown that terms non-analytic in the quark masses do not affect the
shape, and only appear in the normalization constants. Predictive power is
retained including the leading analytic $m_{q}$ operators. With the symmetry
violating corrections we derive useful model-independent relations between $%
\phi _{\pi }^{p,\sigma }$, $\phi _{\eta }^{p,\sigma }$, $\phi
_{K^{+},K^{0}}^{p,\sigma }$, and $\phi _{\bar{K}^{0},K^{-}}^{p,\sigma }$. We
have also commented on the calculation of the moments of these distributions
using lattice method and light-cone sum rules.

\bigskip

\bigskip

We thank Will Detmold, Dan Pirjol and Iain Stewart for useful
comments on the manuscript. Support of NSC and NCTS of ROC is
gratefully acknowledged.

\end{document}